\documentclass[aps,12pt]{revtex4}%
\usepackage{amsfonts}
\usepackage{amsmath}
\usepackage{amssymb}
\usepackage{graphicx}%
\begin{document}
\title{Energy Transport in the Vaidya System}
\author{J. P. Krisch and E. N. Glass}
\affiliation{Department of Physics, University of Michigan, Ann Arbor, MI 48109}
\date{11 March 2005 }

\begin{abstract}
Energy transport mechanisms can be generated by imposing relations between
null tetrad Ricci components. Several kinds of mass and density transport
generated by these relations are studied for the generalized Vaidya system.

\end{abstract}
\maketitle

\section{Introduction}

The Vaidya \cite{Vai53}\ spacetime generalized the Schwarzschild vacuum
solution by allowing mass parameter $m_{0}$ to be a function of retarded
time.\ The extension created a spherically symmetric null fluid
atmosphere.\ Glass and Krisch \cite{GK99a} pointed out that allowing the mass
function to also depend on the radial coordinate created a more complex
atmosphere containing an anisotropic string fluid in addition to Vaidya radiation.

The fluid parameters for the Vaidya atmosphere $\{\rho,$ $p_{r},$ $p_{\perp
}\}$ all depend on time through their relationship to the time dependent mass
function $m(u,r)$; they control the motion of matter through the atmosphere.
The Ricci tensor null tetrad components (i.e. Ricci spinor components),
$\Phi_{11}$ and $\Phi_{22}$, for the generalized Vaidya spacetime also depend
on both spatial and time derivatives of the mass function. If we impose
relations between these components then those relations will, in turn, impose
matter transport in the Vaidya system.\ Conversely, if a\textbf{\ }%
particular\ mode of matter transport\ were assumed for the mass or one of the
fluid parameters, it\textbf{\ }would contain within it a Ricci relation.
Relating Ricci components is a way of imposing and classifying a range of
atmospheric matter transport mechanisms. In this paper we consider the
generalized Vaidya metric \cite{GK99a},\cite{GK98}, and examine how functional
relations between $\Phi_{11}$ and $\Phi_{22}$ lead to mass transport described
by\ the diffusion equation, by the wave equation, and by a dissipative
transport equation of Telegrapher type.

In studying fluid transport, the transport equations are often reducible to
ordinary differential equations by introducing a similarity variable;\ a well
known example of this is the diffusion variable, $\eta\sim r/\sqrt{t}$ ,
suggested by Boltzmann \cite{Ghe88} in 1894. All the transport equations
studied in this paper have similarity solutions. The physical similarity of
matter fields has been discussed by many authors \cite{GK99b},\cite{CH91}%
,\cite{Col97},\cite{YY97}.\textbf{\ }The matter transport considered here is
in the atmosphere around a compact object or an already existing black hole.
By examining the similarity structure of the mass solution to the transport
equations, we are able to relate the function of proportionality between the
Ricci components to the spatial part of a similarity variable.

The paper is structured as follows; In the next section we briefly review the
generalized Vaidya spacetime.\ Matter transport mechanisms and their
similarity structure are discussed in section III. Some explicit examples are
given in section IV. Metric and tetrad details are provided in Appendix A\ 

Our sign conventions are $2A_{\nu;[\alpha\beta]}=A_{\mu}R_{\ \ \nu\alpha\beta
}^{\mu},$ $R_{\mu\nu}=R_{\ \mu\nu\alpha}^{\alpha}$, and metric signature
($+,-,-,-$). Greek indices range over $(0,1,2,3)=(u,r,\vartheta,\varphi)$.
$\dot{m}$ abbreviates $\partial m/\partial u$, $m^{\prime}$ abbreviates
$\partial m/\partial r$, with another prime for each higher derivative.
Overhead carets denote unit vectors. Field equations are $G_{\mu\upsilon
}=-8\pi T_{\mu\upsilon}$.

\section{Generalized Vaidya Spacetime}

The Vaidya metric generalizes vacuum Schwarzschild when $m_{0}\rightarrow
m(u)$.
\begin{equation}
g_{\alpha\beta}^{\text{vad}}dx^{\alpha}dx^{\beta}=A^{\text{vad}}%
du^{2}+2dudr-r^{2}(d\vartheta^{2}+\text{sin}^{2}\vartheta d\varphi^{2})
\label{vad-met}%
\end{equation}
where $A^{\text{vad}}=1-2m(u)/r$. The only non-zero Weyl null tetrad component
is $\Psi_{2}~$=$~-m(u)/r^{3}$. The Ricci tensor, with $l_{\alpha}dx^{\alpha
}=du$, is%
\[
R_{\alpha\beta}^{\text{vad}}=\frac{2\dot{m}}{r^{2}}l_{\alpha}l_{\beta}.
\]

The Vaidya metric has been generalized so that $m(u)\rightarrow m(u,r)$ and
$A^{\text{vad}}\rightarrow A^{\text{gv}}$~$=$~$1$~$-$~$2m(u,r)/r$.%
\begin{equation}
g_{\alpha\beta}^{\text{gv}}dx^{\alpha}dx^{\beta}=A^{\text{gv}}du^{2}%
+2dudr-r^{2}(d\vartheta^{2}+\text{sin}^{2}\vartheta d\varphi^{2}).
\label{gen-vad-met}%
\end{equation}
The metric is Petrov type \textbf{D} with $l_{\mu}$ and $n_{\mu}$ principal
null vectors (details~are~in~Appendix~A).
\begin{subequations}
\label{pnvtet}%
\begin{align}
l_{\mu}dx^{\mu}  &  =du,\label{nteta}\\
n_{\mu}dx^{\mu}  &  =(A^{\text{gv}}/2)du+dr,\label{ntetb}\\
m_{\mu}dx^{\mu}  &  =-(r/\surd2)(d\vartheta+i\ \text{sin}\vartheta d\varphi).
\label{ntetc}%
\end{align}

Basis vectors\ for the generalized Vaidya metric are the unit vector set
($\hat{v}_{\mu},\hat{r}_{\mu},\hat{\vartheta}_{\mu},\hat{\varphi}_{\mu})$ and
the related null set ($l_{\mu},n_{\mu},m_{\mu},\bar{m}_{\mu})$ such that
\end{subequations}
\begin{align}
g_{\mu\nu}^{\text{gv}}  &  =\hat{v}_{\mu}\hat{v}_{\nu}-\hat{r}_{\mu}\hat
{r}_{\nu}-\hat{\vartheta}_{\mu}\hat{\vartheta}_{\nu}-\hat{\varphi}_{\mu}%
\hat{\varphi}_{\nu}\\
&  =l_{\mu}n_{\nu}+n_{\mu}l_{\nu}-m_{\mu}\bar{m}_{\nu}-\bar{m}_{\mu}m_{\nu
}.\nonumber
\end{align}
In terms of the basis vectors, metric $g^{\text{gv}}$ has energy-momentum
\cite{GK99a}
\begin{equation}
-8\pi T_{\mu\nu}^{\text{gv}}=\psi l_{\mu}l_{\nu}+\rho\hat{v}_{\mu}\hat{v}%
_{\nu}+p_{r}\hat{r}_{\mu}\hat{r}_{\nu}+p_{\perp}(\hat{\vartheta}_{\mu}%
\hat{\vartheta}_{\nu}+\hat{\varphi}_{\mu}\hat{\varphi}_{\nu})
\label{en-mom-gv}%
\end{equation}
with components
\begin{subequations}
\label{T-comp}%
\begin{align}
4\pi\psi &  =-\dot{m}/r^{2},\label{psi}\\
4\pi\rho &  =-4\pi p_{r}=m^{\prime}/r^{2},\label{rho}\\
8\pi p_{\perp}  &  =-m^{\prime\prime}/r. \label{pperp}%
\end{align}

The Einstein tensor is computed from $g^{\text{gv}}$ and given by
\end{subequations}
\begin{align}
G_{\mu\upsilon}^{\text{gv}}  &  =-2\Phi_{11}(l_{\mu}n_{\upsilon}+n_{\mu
}l_{\upsilon}+m_{\mu}\bar{m}_{\upsilon}+\bar{m}_{\mu}m_{\upsilon
})\label{ein-gvad}\\
&  -2\Phi_{22}l_{\mu}l_{\upsilon}-(\mathcal{R}/4)g_{\mu\upsilon}^{\text{gv}%
},\nonumber
\end{align}
with components
\begin{subequations}
\label{ein2}%
\begin{align}
\Phi_{11}  &  =(2m^{\prime}-rm^{\prime\prime})/(4r^{2})=2\pi(p_{\perp}%
-p_{r}),\label{ein1a}\\
\Phi_{22}  &  =-\dot{m}/r^{2},\label{ein1b}\\
\mathcal{R}  &  =2(rm^{\prime\prime}+2m^{\prime})/r^{2}=-16\pi(p_{\perp}%
+p_{r}). \label{ein1c}%
\end{align}
An inspection of Eq.(\ref{ein2}) shows that a relation between $\Phi_{11}$ and
$\Phi_{22}$ will generate matter transport.

$S_{\mu\nu}=\frac{1}{2}(R_{\mu\nu}^{\text{gv}}-\frac{1}{4}\mathcal{R}%
^{\text{gv}}g_{\mu\nu})$ is the trace-free Ricci tensor. Its eigenspectrum
determines the Segre type of metric $g_{\mu\nu}^{\text{gv}}$. The
characteristic equation, det[$S-\lambda I$] = $0$, is
\end{subequations}
\begin{equation}
\lambda^{4}-(\Phi_{22})\lambda^{3}-2(\Phi_{11}^{2})\lambda^{2}+(\Phi_{11}%
^{2}\Phi_{22})\lambda+\Phi_{11}^{4}=0, \label{char-eqn}%
\end{equation}
with factors%
\begin{equation}
(\lambda-\Phi_{11})(\lambda+\Phi_{11})(\lambda^{2}-\Phi_{22}\lambda-\Phi
_{11}^{2}). \label{char-fact}%
\end{equation}
This set corresponds to Segre type [2,(11)]. The Ricci relations established
below will not change the Segre type.

\section{GEOMETRY AND\ TRANSPORT OF MATTER}

\subsection{Mass diffusion from fluid relations}

The\ fluid components of the energy-momentum in equation (\ref{T-comp}) depend
on both time and spatial derivatives, so matter transport will
occur.\ Implicit in these equations is the relation%
\[
4\pi r^{2}\dot{\rho}=\partial_{r}\dot{m}.
\]
If one also assumes
\[
\dot{m}=4\pi D_{0}r^{2}\partial_{r}\rho
\]
then both the mass and density obey diffusion equations, the mass diffusing in
a space whose determinant is dual to Vaidya \cite{GK99a}.\textbf{\ }
\begin{subequations}
\label{independ-3}%
\begin{align}
\dot{\rho}  &  =D_{0}\triangledown^{2}\rho,\label{ind1}\\
\dot{m}  &  =D_{0}\nabla^{-2}m \label{ind4}%
\end{align}
with $\nabla^{2}=r^{-2}(\partial/\partial r)r^{2}(\partial/\partial r)$,
$\nabla^{-2}=r^{2}(\partial/\partial r)r^{-2}(\partial/\partial r)$, and
$D_{0}$\ the diffusion constant.\ Equating the non-zero Ricci components will
also produce motion of the mass and density.

\subsection{Mass diffusion from Ricci components}

Consider the quantities ($4r\Phi_{11}$) and ($r^{2}\Phi_{22}$). Equations
(\ref{ein1a}) and (\ref{ein1b}) (with no assumptions) allow the relation
between these quantities to be written as
\end{subequations}
\begin{equation}
\partial_{u}(4r\Phi_{11})=r^{2}\partial_{r}[\frac{\partial_{r}(r^{2}\Phi
_{22})}{r^{2}}]. \label{ricci-rel}%
\end{equation}
If there is a\textbf{\ }general linear relation between $\Phi_{11}$ and
$\Phi_{22}$ such as
\begin{equation}
\Phi_{22}=h_{d}(r)\Phi_{11}. \label{lin-rel}%
\end{equation}
then both Ricci components will evolve diffusively.\textbf{\ }We will show in
the next section that $h_{d}(r)$ is related to a spatial similarity variable
for the diffusion equation. For exact solutions mass is a more physical
quantity and, if we impose this relation and examine mass transport, equations
(\ref{ein1a}) and (\ref{ein1b}) imply
\begin{subequations}
\begin{align}
\Phi_{11}  &  =-(r/4)\partial_{r}(m^{\prime}/r^{2})\label{m-prime}\\
\Phi_{22}  &  =-\dot{m}/r^{2}=-(rh_{d}/4)\partial_{r}(m^{\prime}/r^{2}).
\label{m-dot}%
\end{align}
The linear relation (\ref{lin-rel}) together with (\ref{m-dot}) yields
\end{subequations}
\begin{equation}
\dot{m}=(r^{3}h_{d}/4)\partial_{r}(m^{\prime}/r^{2}). \label{m-diffu}%
\end{equation}
and so the mass will also move diffusively.\ 

The homogeneous solution of Eq.(\ref{m-diffu}) is%
\begin{align}
m  &  =r^{3}c_{1}(u)+c_{2}(u),\text{ \ }\dot{c}_{2}+r^{3}\dot{c}_{1}=0\\
8\pi\rho &  =-8\pi p_{r}=6c_{1}(u)\nonumber\\
p_{r}  &  =p_{\perp}.\nonumber
\end{align}
Equation $\dot{c}_{2}+r^{3}\dot{c}_{1}=0$ is satisfied only for $c_{1}$ and
$c_{2}$ both constant.

\subsection{Mass transport by wave motion}

The basic relation between the Ricci components, Eq.(\ref{ricci-rel}), is%
\[
\partial_{u}(4r\Phi_{11})=r^{2}\partial_{r}[\frac{\partial_{r}(r^{2}\Phi
_{22})}{r^{2}}].
\]
If one assumes the Ricci components are related by%
\begin{equation}
\partial_{u}\Phi_{22}=h_{w}(r)\Phi_{11}%
\end{equation}
then $r^{2}\Phi_{22}$ will evolve as a wave equation solution.\ Focusing again
on the mass behavior, Eq.(\ref{ein1a}) for $\Phi_{11}$ and Eq.(\ref{ein1b})
for $\Phi_{22}$ provide a wave equation for the mass:
\begin{equation}
\ddot{m}-(h_{w}/4)r^{3}\triangledown^{-2}m=0\text{ }\nonumber
\end{equation}
or
\begin{equation}
\ddot{m}-(h_{w}/4)r^{5}\partial_{r}(r^{-2}\partial_{r}m)=0. \label{mass-wave}%
\end{equation}

Classical wave motion provides shape preserving traveling solutions to the
wave equations for functions with argument $(kr-\omega t).$ Therefore,
consider a variable of the form\ $\eta=R(r)-T(u).$ The wave equation becomes%
\begin{equation}
\lbrack\dot{T}^{2}-(h_{w}/4)r^{3}(R^{\prime})^{2}]m_{\eta\eta}-[\ddot
{T}+(h_{w}/4)(r^{3}R^{\prime\prime}-2r^{2}R^{\prime})]m_{\eta}=0.
\end{equation}
The transport equation is%
\begin{align*}
\alpha m_{\eta\eta}-\beta m_{\eta}  &  =0\\
\dot{T}^{2}-(h_{w}/4)r^{3}(R^{\prime})^{2}  &  =\alpha\\
\ddot{T}+(h_{w}/4)(r^{3}R^{\prime\prime}-2r^{2}R^{\prime})  &  =\beta.
\end{align*}
The simplest solutions require $\alpha$ and $\beta$ to be separately zero.
These are
\begin{equation}
h_{w}=\frac{4c_{1}}{r^{3}(R^{\prime})^{2}},~~R=c_{2}r^{3}+c_{3},~~T=T_{0}%
u+T_{1},
\end{equation}
and, as in diffusive transport, the proportionality function between the Ricci
components is related to the similarity variable. The mass is given by any
function of argument $\eta$, $m=F(\eta).$

\subsection{Dissipative transport}

It is clear that the mode of matter transport depends on the relation chosen
between $\Phi_{22}\ $and$\ \Phi_{11}.$ Dissipative waves obey a Telegrapher
equation \cite{Ros93}, combining both wave and diffusive elements.
\begin{equation}
\ddot{\chi}-c_{0}^{2}\triangledown^{2}\chi+c_{1}\dot{\chi}=0
\end{equation}
where $\triangledown^{2}\chi=(1/r^{2})\partial_{r}(r^{2}\partial_{r}\chi).$
The relationship%
\begin{equation}
(\partial_{u}+\alpha_{0})\Phi_{22}+h_{t}(r)\Phi_{11}=0 \label{phi11-phi22}%
\end{equation}
will generate an equation of Telegrapher form for mass%
\begin{equation}
\ddot{m}-(h_{t}/4)r\triangledown^{-2}m+\alpha_{0}\dot{m}=0.
\end{equation}
This form of causal dissipative transport provides a richer tool for modeling
than a pure diffusion equation, since it avoids the infinite propagation
speeds associated with parabolic diffusion equations \cite{JP89}. Causal
transport has recently been discussed by Herrera and Santos \cite{HS04}.\ As
before, the function of proportionality $h_{t}$ is related to a similarity variable.

\subsection{Similarity and $h_{d}$}

To see the relation of the proportionality function to the similarity
structure of the diffusion equation, assume a mass solution and similarity
variable with forms
\begin{align*}
m-m_{0}  &  =T(u)F(\eta)\\
\eta &  =R(r)/L(u).
\end{align*}
Rewriting the mass diffusion equation (\ref{m-diffu}), with $F_{\eta}%
=dF/d\eta$ results in
\begin{equation}
F_{\eta\eta}+F_{\eta}\left[  L\{\frac{R^{\prime\prime}}{(R^{\prime})^{2}%
}-\frac{2}{rR^{\prime}}\}+\frac{4R}{h_{d}r(R^{\prime})^{2}}\dot{L}\right]
-\left[  \frac{4}{h_{d}r(R^{\prime})^{2}}L^{2}\frac{\dot{T}}{T}\right]  F=0.
\label{simm-diffu1}%
\end{equation}
This diffusion equation should have similarity form \cite{Ghe88}
\[
F_{\eta\eta}+2\alpha\eta F_{\eta}-\beta F=0.
\]
We look for solutions with the same similarity form, and so the coefficients
impose constraints.\ The coefficient of $F$ relates $\beta$ and $h_{d}(r)$%
\begin{equation}
\beta=\frac{4}{h_{d}r(R^{\prime})^{2}}L^{2}\frac{\dot{T}}{T}. \label{beta-eqn}%
\end{equation}
Removing r-dependence provides the relation between $R(r)$ and $h_{d}(r):$%
\begin{equation}
h_{d}=\frac{4c_{0}}{r(R^{\prime})^{2}}.
\end{equation}
With this $h_{d}$ constraint, the u-dependent part of the coefficient $\beta$
implies
\begin{equation}
L^{2}\frac{\dot{T}}{T}=\beta c_{0}. \label{beta-c0-const}%
\end{equation}
This allows Eq.(\ref{simm-diffu1}) to be written as%
\[
F_{\eta\eta}+F_{\eta}\left[  L\{\frac{R^{\prime\prime}}{(R^{\prime})^{2}%
}-\frac{2}{rR^{\prime}}\}+(R/c_{0})\dot{L}\right]  -\beta F=0.
\]
The coefficient of $F_{\eta}$ requires
\begin{equation}
2\alpha\eta=2\alpha(R/L)=L[\frac{R^{\prime\prime}}{(R^{\prime})^{2}}-\frac
{2}{rR^{\prime}}]+(R/c_{0})\dot{L}. \label{beta-l-eqn}%
\end{equation}
Solutions of this equation depend on the value of $\alpha$. Some example
solutions will be given in Section IV. We note here that the choice
$R=R_{0}r^{3}+R_{1}$, which solves%
\begin{equation}
R^{\prime\prime}/R^{\prime}-2/r=0, \label{spatial-scale}%
\end{equation}
provides an analog of the Boltzman similarity variable.\ For this choice
Eq.(\ref{beta-l-eqn}) yields
\[
L^{2}=4c_{0}\alpha u+L_{0}^{2}%
\]
with similarity variable $R(r)/L(u)$%
\[
\eta=\frac{R_{0}r^{3}+R_{1}}{\sqrt{4c_{0}\alpha u+L_{0}^{2}}}%
\]
\ and proportionality function%
\[
h_{d}=\frac{4c_{0}}{r^{5}R_{0}^{2}}.
\]
If $c_{0}=D_{0}R_{0}^{2}$ then $h_{d}=4D_{0}/r^{5}$. This choice yields a
simple diffusion equation for mass.\ 

\subsection{Similarity and h$_{t}$}

Assuming a mass and similarity variable of the form%
\[
m-m_{0}=T(u)F(\eta),\text{ \ \ \ }\eta=R(r)/L(u)
\]
we find
\begin{align*}
\left[  \frac{\dot{L}^{2}}{L^{2}}\frac{R^{2}}{T}-(\frac{h_{t}}{4})r(R^{\prime
})^{2}\right]  F_{\eta\eta}  & \\
-\left[  R(2\dot{L}\frac{\dot{T}}{T}+\ddot{L}-2\frac{\dot{L}^{2}}{L}%
+\alpha_{0}\dot{L})+(\frac{h_{t}}{4})L(rR^{\prime\prime}-2R^{\prime})\right]
F_{\eta}  & \\
+\left[  L^{2}(\frac{\ddot{T}}{T}+\alpha_{0}\frac{\dot{T}}{T})\right]  F=0  &
\end{align*}
The relationship of $h_{t}$\ to the scaling variable in this similarity
equation depends on the form of the time scaling function. \ The example that
we shall use in the next section has $L=L_{0}$, and for this choice the
similarity equation becomes%
\begin{equation}
-\frac{h_{t}}{4}r(R^{\prime})^{2}F_{\eta\eta}+L_{0}\frac{h_{t}}{4}(2R^{\prime
}-rR^{\prime\prime})F_{\eta}+L_{0}^{2}(\frac{\ddot{T}}{T}+\alpha_{0}\frac
{\dot{T}}{T})F=0.
\end{equation}
For this choice of $L$, we see the diffusive relation between the
proportionality function and the spatial scaling function again emerges as\ \
\begin{equation}
h_{t}=\frac{4c_{0}}{r(R^{\prime})^{2}}.
\end{equation}

Telegrapher transport could allow many other relations to be imposed,
reflecting the richer solution structure of this transport mechanism.\ In the
next section we give some examples. \ 

\section{TRANSPORT\textbf{\ EXAMPLES}}

\subsection{Similarity Solutions}

As an example of the mass transport solutions based on similarity, we consider
a single scenario to which we apply all three transport mechanisms. \ The
physical setting considered is a compact object of mass $m(u,\eta)$\ with an
atmosphere in which transport is occurring.\ We use solutions where the scale
variable $R$\ and proportionality function $h(r)$\ are the same for all
transports.\textbf{\ }%
\begin{equation}
R=\frac{R_{0}}{3}r^{3},~~L=L_{0},~~h(r)=\frac{4c_{0}}{R_{0}^{2}r^{5}}.
\end{equation}
In simple diffusion problems, the choice $L=L_{0}$\ is applied to bounded
systems where $L_{0}$\ can be identified with a natural physical scale. \ For
the Vaidya black hole, the Schwarzschild radius provides a physical distance
scale and we could identify $L_{0}$\ with an associated distance.

\subsubsection{Diffusive transport}

For diffusive transport with the choices above, the equations describing the
evolution of the mass function are
\begin{equation}
F_{\eta\eta}-\beta F=0,\text{\ \ \ \ \ \ }\dot{T}/T=\beta c_{0}/L_{0}^{2}.
\end{equation}
The mass function, with time parameter $\tau_{0}^{-1}=\beta D_{0}R_{0}%
^{2}/L_{0}^{2}$, will be given by
\begin{align}
m  &  =m_{0}+T_{0}e^{\ u/\tau_{0}}~F(\eta)\\
F(\eta)  &  =F_{0}\sin(\sqrt{\left\vert \beta\right\vert }\eta+\delta),\text{
\ }\beta<0\nonumber\\
&  =F_{0}\sinh(\sqrt{\beta}\eta+\delta),\text{ \ }\beta>0.\nonumber
\end{align}
For $\beta<0$\ , the atmosphere is decaying in time as the Vaidya photons
carry energy out of the system, while for $\beta>0,$\ mass is accreting. From
the field equations, the density of the atmosphere is described by%
\begin{align}
\rho &  =\rho_{0}e^{-u/\tau_{0}}\cos(\sqrt{\left\vert \beta\right\vert }%
\eta+\delta)\\
&  =\rho_{0}e^{u/\tau_{0}}\cosh(\sqrt{\beta}\eta+\delta)\nonumber
\end{align}
where we have identified the initial atmospheric density\ and a time constant
for the accretion or decay
\begin{equation}
\rho_{0}=\frac{T_{0}F_{0}\sqrt{\left\vert \beta\right\vert }}{2\pi L_{0}}.
\end{equation}

The similarity variable in this example is simply a distance coordinate. The
decaying solution could describe a bounded atmosphere whose density decreases
with distance away from the surface. There is zero radial pressure outer
boundary described by $\sqrt{\left\vert \beta\right\vert }\eta_{B}+\delta
=\pi/2.$ In the decaying solutions, the atmosphere will go asymptotically to
zero leaving a compact object of mass $m_{0}.$\ Because of the outgoing Vaidya
radiation, there is no vacuum match until the atmosphere is gone. For the
accreting solution, there is no zero radial pressure boundary. Since the
accreting mass is entering the atmosphere from the exterior, this is expected.
The density profile depends strongly on the phase $\delta.$ For $\delta=0,$
the density increases as one looks upward from the core surface.\ For non-zero
$\delta$, the density decreases going up from the surface, reaching a minimum
value which might be identified with the boundary between the atmosphere and
the source of the accreting mass. \ 

\subsubsection{Telegrapher transport\ }

The equations for Telegrapher transport with the scaling functions chosen are
\begin{align*}
F_{\eta\eta}-\beta F  &  =0\\
\frac{L_{0}^{2}}{c_{0}}(\frac{\ddot{T}}{T}+\alpha_{0}\frac{\dot{T}}{T})  &
=\beta
\end{align*}
Telegrapher transport has a wider set of solutions for the scaling choice than
pure diffusive transport.\ One solution, using $\beta=0$,\ is%
\begin{align}
T  &  =T_{0}e^{-\alpha_{0}u}\\
F  &  =F_{0}\eta\nonumber\\
m  &  =m_{0}+\rho_{1}r^{3}e^{-\alpha_{0}u},\text{ \ }\rho_{1}=T_{0}F_{0}%
/L_{0}.\nonumber
\end{align}
The atmospheric density is related to the mass through the field equations by
$\rho=m^{\prime}/(4\pi r^{2})$. \ The density for this case does not vary with
radial distance from the black hole but does reflect the atmospheric decay,
approaching zero as the atmosphere vanishes and the mass becomes $m_{0}%
$:\textbf{\ \ }%
\begin{equation}
\rho=\rho_{0}e^{-\alpha_{0}u},\text{ \ }\rho_{0}=3R_{0}\rho_{1}/4\pi.
\end{equation}

The $\beta\neq0$\ solutions are similar to the diffusive solutions%
\begin{align}
T  &  =T_{0}e^{\gamma u},\text{ \ }\beta=(L_{0}^{2}/c_{1})(\gamma^{2}%
+\alpha_{0}\gamma)\\
F  &  =F_{0}\sin(\sqrt{\left\vert \beta\right\vert }\eta+\delta),\text{
\ }\beta<0\nonumber\\
F  &  =F_{0}\sinh(\sqrt{\beta}\eta+\delta),\text{ \ }\beta>0\nonumber\\
m  &  =m_{0}+T(u)F(\eta).\nonumber
\end{align}

\subsection{A diffusively evaporating atmosphere}

\paragraph{The boundary behavior $\blacklozenge$}

In this example, we consider a compact object whose atmosphere is diffusively
evaporating. An exact mass solution, given in \cite{GK99a}, is$\ \ $%
\begin{equation}
m=m_{0}+(4\pi/3)r^{3}\rho_{0}-(4\pi/3)k_{2}(r^{5}/10+\mathfrak{D}_{0}ur^{3}).
\end{equation}
For this example the proportionality function follows directly from the
diffusion equation\textbf{\ }%
\[
h_{d}=4D_{0}/r.
\]
The atmospheric density associated with this mass is%
\begin{equation}
\rho=\rho_{0}-k_{2}(r^{2}/6+\mathfrak{D}_{0}u).
\end{equation}

A boundary can be defined by requiring the radial pressure (and density, from
the equation of state) to be zero. Solving the density equation for a boundary
radius gives the boundary as a function of retarded time
\[
R_{b}^{2}=\frac{6}{k_{2}}(\rho_{0}-k_{2}\mathfrak{D}_{0}u).
\]
Let the boundary move inward as the atmosphere evaporates until the atmosphere
is gone and the radius is at the compact object boundary, $R_{c}$. This
happens in time $u_{0}$, thus%
\begin{equation}
R_{c}^{2}=\frac{6}{k_{2}}(\rho_{0}-k_{2}\mathfrak{D}_{0}u_{0}). \label{e2}%
\end{equation}
It follows that%
\begin{equation}
R_{b}^{2}(u)=R_{c}^{2}+6\mathfrak{D}_{0}(u_{0}-u). \label{e1-e2}%
\end{equation}
The bounding surface of the core is parameterized as $R_{c}=2m_{0}\alpha.$
Substituting into (\ref{e1-e2}) provides%
\begin{equation}
R_{b}^{2}(u)=4m_{0}^{2}\alpha^{2}+6\mathfrak{D}_{0}(u_{0}-u). \label{e3}%
\end{equation}
The mass function places a constraint on the parameters. From Eq.(\ref{e2}) at
time $u_{0}$
\begin{equation}
(\alpha-1)m_{0}=\frac{128}{45}k_{2}\pi m_{0}^{5}\alpha^{5}. \label{e4}%
\end{equation}
Using Eqs.(\ref{e2}) and (\ref{e4}), the time for the evaporation process to
complete is%
\begin{equation}
\mathfrak{D}_{0}u_{0}=\frac{2}{3}\alpha^{2}m_{0}^{2}\left[  \frac{64\pi
\rho_{0}m_{0}^{2}\alpha^{3}}{15(\alpha-1)}-1\right]  . \label{e5}%
\end{equation}
The density and mass evolve as%
\begin{align}
\rho &  =\frac{k_{2}}{6}(R_{c}^{2}-R_{b}^{2})+k_{2}\mathfrak{D}_{0}(u_{0}-u)\\
m  &  =m_{0}+\frac{4\pi k_{2}}{45}R_{b}^{5}+\frac{4\pi}{3}k_{2}R_{b}%
^{3}\left[  \frac{R_{c}^{2}-R_{b}^{2}}{6}+\mathfrak{D}_{0}(u_{0}-u)\right]  .
\end{align}

\paragraph{Extremal case $\blacklozenge$}

The time at which the evaporation of the atmosphere is complete has an
extremal value.\ Extremising Eq.(\ref{e5}) one finds
\[
\frac{2(\alpha-1)^{2}}{\alpha^{3}(4\alpha-5)}=\frac{64\pi\rho_{0}m_{0}^{2}%
}{15}.
\]
The extremal time for the evaporation is
\begin{equation}
\mathfrak{D}_{0}u_{0}=\frac{2}{3}\alpha^{2}m_{0}^{2}[\frac{3-2\alpha}%
{4\alpha-5}]. \label{extreme-D}%
\end{equation}
The second derivative shows this is a minimum. For positive times, we require
\[
1.25<\alpha<1.5
\]

\paragraph{Time estimates $\blacklozenge$}

Take $m_{0}$ to be a solar size : $m_{0}\sim1.5\times10^{3}m.$ The diffusivity
constant $\mathfrak{D}_{0}$ is related to the constancy of the jump distance
($L$) and frequency ($f$) \cite{Ghe88}. $\mathfrak{D}_{0}$ can be estimated
as
\[
\mathfrak{D}_{0}\sim L^{2}f.
\]
For example, if the Debye frequency in a solid is the same as the jump
frequency, we would have $f\sim10^{13}Hz$ and the jump distance could be of
the order of an Angstrom, so that an estimate for the diffusivity is%
\[
\mathfrak{D}_{0}\sim10^{-7}m^{2}/s.
\]
Using these estimates, the time for the atmosphere to vanish by diffusion is
roughly%
\begin{align*}
u_{0}  &  =10^{7}(\frac{2}{3})\alpha^{2}\times2.25\times10^{6}[\frac
{3-2\alpha}{4\alpha-5}]\\
&  \sim1.5\times10^{13}[\frac{3-2\alpha}{4\alpha-5}]\alpha^{2}%
\end{align*}
For $\alpha=1.4$, the time is about $10^{13}$ sec $\sim$ $10^{5}$ years.\ The
diffusivity is much larger than is normally measured since not all atomic
oscillations will have an associated jump. A smaller diffusivity would
increase the time.

\section{DISCUSSION}

In this paper we have examined some of the atmospheric matter transport
mechanisms introduced by imposing a relation between Ricci components of the
generalized Vaidya spacetime. We found that diffusive, wave, and general
Telegrapher-type transport can all follow from such a relation.\textbf{\ }%
Imposing a Ricci relation is a way of unifying matter transport
mechanisms;\ if one started with a particular transport mechanism then a Ricci
relation would emerge.\ 

Examples of the matter transport mechanisms were given.\ All of the exact
solutions describe atmospheres that are either accreting or decaying. The
first two examples had the same similarity function but the density profiles
in each solution were very different. The first Telegrapher solution describes
a core object with a decaying atmospheric density varying only with time, and
whose tangential and radial stress are all spatially constant tensions.\ In
the pure diffusive solution, the densities vary with both distance from the
core surface and time. The decaying atmospheres could be bounded by an
exterior fluid of Vaidya radiation. The accreting solutions were surrounded by
the source of accreting matter and Vaidya radiation. The third example, with a
different proportionality function, examined a diffusing atmosphere around a
core object of radius $R_{c}$\ in the range \ $2.5m_{0}<R_{c}<3m_{0}$. \ For
an $R_{c}=2.8m_{0}$ object with a diffusivity based on a solid Debye frequency
there was a minimum evaporation time of about $10^{5}$ years. Jump frequencies
based on the actual atmospheric fluid would increase this time.\ Apart from
the actual size of the evaporation time for a specific object, the model
predicts that atmospheres around smaller core objects, near the lower end of
the range, will take very much longer to diffusively evaporate than the
atmospheres of larger core objects. This can be understood by looking at the
initial size of the atmosphere. Using Eq.(\ref{e3}) at $u=0$,
Eq.(\ref{extreme-D}), and the parametrized boundary surface $R_{c}%
=2m_{0}\alpha$, the initial atmospheric radius is%
\[
R_{b}^{2}\mid_{u=0}\text{ }=2R_{c}^{2}\left[  \frac{\alpha-1}{4\alpha
-5}\right]  ,\text{ \ }1.25<\alpha<1.5.
\]
The smaller core objects have larger atmospheric envelopes, taking longer to
evaporate. The smaller core has an extended atmosphere because its
gravitational field is not strong enough to hold the atmosphere compactly.\ \ 

Some additional insights into the meaning of the proportionality function can
be seen for diffusive and wave transport by writing the transport equations in
terms of variable $w=r^{3}.$\ The diffusive and wave transport equations
become\textbf{\ }%
\begin{align*}
\dot{m}  &  =(\frac{9r^{5}h_{d}}{4})m_{ww}\\
\ddot{m}  &  =(\frac{9r^{5}h_{w}}{4})m_{ww}%
\end{align*}
In each case we can find the functional form of $h(r)$\ that will produce the
simplest transport and can relate that value to the diffusion constant or the
wave velocity.\ Our example solutions include this simple case and a case
where the mass and density evolution are more complex.

The choice $h_{d}(r)=4D_{0}/r^{5}$ yields a simple (u,w) diffusion equation
for mass. $\ $From the Ricci relations and the field equations we see that
\[
\dot{m}/r^{2}=2\pi h_{d}(p_{r}-p_{\perp})
\]
indicating that, with the Ricci relations, mass transport is driven by the
pressure anisotropy. This also illustrates a drawback of the mass transport
ansatz.\ $\Phi_{11}$ is certainly zero for isotropic pressures with a string
equation of state [Eq.(\ref{ein1a}) with $p_{\perp}=p_{r}$].\ The assumed
relation,\ $\Phi_{22}=h_{d}(r)\Phi_{11}$, then says that there is no time
variation in the mass for isotropic pressures.\ In general, when the two Ricci
components are not related, the mass can vary with time, even in the case of
isotropic pressures.

We list\ the assumptions that lead to matter transport.\ For diffusion
the\ specific choice $h_{d}=4D_{0}/r^{5}$ is used.

$%
\begin{array}
[c]{ll}
& \\
\underline{\text{Assumption}} & \underline{\text{Transport}}\\
\dot{m}=4\pi D_{0}r^{2}\partial_{r}\rho & \dot{m}-D_{0}\nabla^{-2}m=0,\text{
\ }\dot{\rho}-D_{0}\triangledown^{2}\rho=0\\
\Phi_{22}=(4D_{0}/r^{5})\Phi_{11} & \dot{m}-D_{0}\nabla^{-2}m=0\\
\partial_{u}\Phi_{22}=h_{w}(r)\Phi_{11} & \ddot{m}-(h_{w}/4)r^{3}%
\triangledown^{-2}m=0\\
(\partial_{u}+\alpha_{0})\Phi_{22}+h_{t}(r)\Phi_{11}=0~~~ & \ddot{m}%
-(h_{t}/4)r\triangledown^{-2}m+\alpha_{0}\dot{m}=0\\
&
\end{array}
$

We have seen that imposing a Ricci relation provides a broad arena to
investigate a range of atmospheric transport processes in the generalized
Vaidya spacetime and is a rich source of new analytic mass solutions.\ The
mass solutions that we presented focused on the growth or depletion of an
atmosphere around a central object.\ They can be used to describe the
behavior, for example, of isolated black hole atmospheres but also offer
simple models of galaxies with a massive black hole at the center.\ The
relationships we imposed on the Ricci components were investigated in terms of
mass transport although the fundamental relationships described the evolution
of the Ricci components themselves. The Ricci evolution is an interesting
avenue for further investigation as they are input functions for the Riemann
invariants.\ The evolution of the invariants and their syzigies will be
discussed elsewhere. \ \ 

\appendix

\section{Generalized Vaidya Principal Null Frame}

The principal null frame Eq.(\ref{pnvtet}) of the Petrov type \textbf{D}
metric $g_{\mu\nu}^{\text{gv}}$ obeys
\begin{subequations}
\begin{align}
l_{\mu;\nu}  &  =(A_{\text{gv}}^{^{\prime}}/2)l_{\mu}l_{\nu}-(1/r)(m_{\mu}%
\bar{m}_{\nu}+\bar{m}_{\mu}m_{\nu}),\label{l_covd}\\
n_{\mu;\nu}  &  =-(A_{\text{gv}}^{^{\prime}}/2)n_{\mu}l_{\nu}+(A^{\text{gv}%
}/2r)(m_{\mu}\bar{m}_{\nu}+\bar{m}_{\mu}m_{\nu}),\label{n_covd}\\
m_{\mu;\nu}  &  =(A^{\text{gv}}/2r)l_{\mu}m_{\nu}-(1/r)n_{\mu}m_{\nu
}+(\text{cot}\vartheta/\sqrt{2}r)(m_{\mu}m_{\nu}-m_{\mu}\bar{m}_{\nu}).
\label{m_covd}%
\end{align}
with both principal null vectors $l_{\mu}$ and $n_{\mu}$ geodesic.

For tetrad $\{\hat{v},\hat{r},\hat{\vartheta},\hat{\varphi}\}$ and metric
$g_{\mu\nu}^{\text{gv}}=\hat{v}_{\mu}\hat{v}_{\nu}-\hat{r}_{\mu}\hat{r}_{\nu
}-\hat{\vartheta}_{\mu}\hat{\vartheta}_{\nu}-\hat{\varphi}_{\mu}\hat{\varphi
}_{\nu}$, the basis vectors are related by
\end{subequations}
\begin{subequations}
\label{vtet}%
\begin{align}
\hat{v}_{\mu}dx^{\mu}  &  =A_{\text{gv}}^{1/2}du+A_{\text{gv}}^{-1/2}%
dr=A_{\text{gv}}^{-1/2}[n_{\mu}+(A_{\text{gv}}/2)l_{\mu}]dx^{\mu
},\label{vteta}\\
\hat{r}_{\mu}dx^{\mu}  &  =A_{\text{gv}}^{-1/2}dr=A_{\text{gv}}^{-1/2}[n_{\mu
}-(A_{\text{gv}}/2)l_{\mu}]dx^{\mu},\label{vtetb}\\
\hat{\vartheta}_{\mu}dx^{\mu}  &  =rd\vartheta=(1/\surd2)(m_{\mu}+\bar{m}%
_{\mu})dx^{\mu},\label{vtetc}\\
\hat{\varphi}_{\mu}dx^{\mu}  &  =r\text{sin}\vartheta d\varphi=-(i/\surd
2)(m_{\mu}-\bar{m}_{\mu})dx^{\mu}. \label{vtetd}%
\end{align}
The kinematics of the $\hat{v}$ flow, acceleration, expansion, and shear, are
described by
\end{subequations}
\begin{equation}
\hat{v}_{\ ;\nu}^{\mu}=a^{\mu}\hat{v}_{\nu}+\sigma_{\ \nu}^{\mu}%
-(\Theta/3)(\hat{r}^{\mu}\hat{r}_{\nu}+\hat{\vartheta}^{\mu}\hat{\vartheta
}_{\nu}+\hat{\varphi}^{\mu}\hat{\varphi}_{\nu}), \label{vflow}%
\end{equation}
where
\begin{subequations}
\begin{align}
a^{\mu}  &  =[\dot{m}/r+A_{\text{gv}}\partial_{r}(m/r)]A_{\text{gv}}%
^{-3/2}\hat{r}^{\mu},\label{va}\\
\sigma_{\ \nu}^{\mu}  &  =(\Theta/3)(-2\hat{r}^{\mu}\hat{r}_{\nu}%
+\hat{\vartheta}^{\mu}\hat{\vartheta}_{\nu}+\hat{\varphi}^{\mu}\hat{\varphi
}_{\nu}),\label{vb}\\
\Theta &  =(\dot{m}/r)A_{\text{gv}}^{-3/2}. \label{vc}%
\end{align}

Spherical symmetry allows the function $m(u,r)$ to be identified as the mass
within two-surfaces of constant $u$ and $r$, and invariantly defined from the
sectional curvature of those surfaces:
\end{subequations}
\begin{equation}
-2m/r^{3}=R_{\alpha\beta\mu\nu}\hat{\vartheta}^{\alpha}\hat{\varphi}^{\beta
}\hat{\vartheta}^{\mu}\hat{\varphi}^{\nu}.
\end{equation}


\begin{thebibliography}{99}                                                                                               %


\bibitem {Vai53}P.C. Vaidya, Nature \textbf{171,} 260 (1953).\ 

\bibitem {GK99a}E.N. Glass and J.P. Krisch, Class. Quantum Grav. \textbf{16},
1175 (1999).

\bibitem {GK98}E.N. Glass and J. P. Krisch, Phys. Rev. D \textbf{57}, R5945 (1998).

\bibitem {Ghe88}R. Ghez, \textit{A Primer of Diffusion Problems}, (John Wiley,
New York, 1988).

\bibitem {GK99b}E.N. Glass and J.P. Krisch, J. Math. Phys. \textbf{40}, 4056 (1999).

\bibitem {CH91}B. Carter and R.N. Henriksen, J. Math. Phys, \textbf{32}, 2580 (1991).

\bibitem {Col97}A.A. Coley, Class. Quantum Grav. \textbf{14}, 87 (1997).

\bibitem {YY97}I. Yavuz and I. Yilmaz, Gen. Rel. Grav. \textbf{29}, 1295 (1997).

\bibitem {Ros93}P. Rosenau, Phys. Rev. E \textbf{48}, R655 (1993).

\bibitem {JP89}D.D. Joseph and L. Preziosi, Rev. Mod. Phys. \textbf{61}, 41 (1989).

\bibitem {HS04}L. Herrera and N.O. Santos, "Dynamics of dissipative
gravitational collapse", gr-qc/0410014 (2004).
\end{thebibliography}
\end{document}